\documentclass[]{aastex631}

\usepackage{amsmath}
\usepackage{amsfonts}
\usepackage{amssymb}
\usepackage[normalem]{ulem}
\usepackage{soul}

\sloppypar
\definecolor{g2}{rgb}{0.0, 0.5, 0.0}

\newcommand{\revision}[1]{#1}

\newcommand{\cca}{Center for Computational Astrophysics, Flatiron Institute, New York, NY 10010, USA}
\newcommand{\pihpi}{$\Pi \mathbf{H} \Pi \,$}
\newcommand{\gradnotation}{$\nabla p$}
\graphicspath{{./}{figs/}{figs/3D/}{figs/slicing/}{figs/viz_data/}}

\shorttitle{The emptiness inside: Finding gaps with geometric data analysis}
\shortauthors{Contardo et al.}

\begin{document}
\title{
The emptiness inside: Finding gaps, valleys, and lacunae with geometric data analysis.}

\correspondingauthor{Gabriella Contardo ~\\A python implementation of the methods presented here is available at \url{https://github.com/contardog/FindTheGap}}

\author[0000-0002-3011-4784]{Gabriella Contardo}
\affil{\cca} 

\author[0000-0003-2866-9403]{David W. Hogg}
\affil{\cca}
\affil{Center for Cosmology and Particle Physics, Department of Physics, New York University}
\affil{Max-Planck-Institut fur Astronomie, Heidelberg}

\author[0000-0001-8917-1532 ]{Jason A. S. Hunt}
\affil{\cca}

\author[0000-0003-4797-7030]{Joshua E. G. Peek}
\affil{Space Telescope Science Institute, 3700 San Martin Drive, Baltimore, MD 21218, USA}
\affil{Department of Physics \& Astronomy, Johns Hopkins University, Baltimore, MD 21218, USA}

\author[0000-0002-4485-306X]{Yen-Chi Chen}
\affil{Department of Statistics, University of Washington}

\begin{abstract}\noindent
Discoveries of gaps in data have been important in astrophysics. For example, there are kinematic gaps opened by resonances in dynamical systems, or exoplanets of a certain radius that are empirically rare. A \emph{gap} in a data set is a kind of anomaly, but in an unusual sense: Instead of being a single \emph{outlier} data point, situated far from other data points, it is a region of the space, or a set of points, that is anomalous compared to its surroundings. Gaps are both interesting and hard to find and characterize, especially when they have non-trivial shapes. \revision{We present in this paper a statistic that can be used to estimate the (local) ``gappiness'' of a point in the data space. It uses the gradient and Hessian of the density estimate (and thus requires a twice-differentiable density estimator). This statistic can be computed at (almost) any point in the space and does not rely on optimization; it allows to highlight under-dense regions of any dimensionality and shape in a general and efficient way. }
We illustrate our method on the velocity distribution of nearby stars in the Milky Way disk plane, which exhibits gaps that could originate from different processes. Identifying and characterizing those gaps could help determine their origins. 
\revision{We provide in an Appendix implementation notes and additional considerations for finding under-densities in data, using critical points and the properties of the Hessian of the density.} 

\end{abstract}


\section{Introduction}

Hypothesis generation is a fundamentally unsolved problem in astronomy, and even more so in the era of large data sets.
We have powerful tools for testing existing hypotheses against the big data sets of the 2020s, but we do not have a clear path for wholly new, unanticipated discoveries when the data scales reach the petabyte regime and data cannot easily be inspected by eye.
Some significant progress has been made in finding rare anomaly objects in astronomical data sets, including using citizen science approaches (e.g. Hanny's Voorverp object \citep{lintott2009galaxy}, green peas galaxies \citep{izotov2011green}, Boyajian's Star \citep{boyajian2016planet}) and machine learning approaches (\cite{baron2017weirdest, margalef2020detecting, martinez2021method, storey2021anomaly})  and simply pure serendipity, as pointed out in \cite{harwit2019cosmic}.
In this work however, we focus on automated detection of the \emph{paucity} of sources, in the form of gaps in the density distribution. Hence, instead of looking for anomalous, rare, (that is, `out-lying') objects, we are looking for regions of the space (or sets of objects) that have anomalously low density (that is, in comparison to their immediate surroundings).

Gaps (local under-densities in the distribution) often have significance to our understanding of the universe, or represent important discoveries. For instance, the ``green valley'' \citep{strateva2001color, baldry2004quantifying, baldry2006galaxy, schawinski2014green} is a gap in the relationship between galaxy mass and specific star formation. In this case, it is thought that the gap is caused by fast evolution of galaxies across the gap from the star-forming sequence to the passively-evolving quiescent galaxies.
Another gap has been discovered in the \textsl{Gaia} DR2 color-magnitude diagram of stars in \cite{jao2018gap}. It is thought to represent either $^3$He instabilities or transitions to full convection in M-dwarfs \citep{feiden2021gaia}. `Gaps' in stellar population have also been of interest, such as the rotational gap as defined by \cite{barnes2010simple}, which was observed in \textsl{Kepler} data by \cite{mcquillan2013measuring} and in K2 in \cite{gordon2021stellar}. In exoplanets population, the `radius valley' has also been under scrutiny, and analysis of this `gap' in the radius of exoplanets as a function of other properties (such as orbital period, stellar mass, and stellar age \citep{fulton2017california, gupta2020signatures, berger2020gaia, david2021evolution}) have been conducted to further understand its underlying mechanisms and causes. 
 And the gaps in Saturn's rings \citep{holberg1982identification} and the asteroid belt \citep{dermott1983nature} reveal important dynamical resonances.
 Similar kinds of gaps appear in velocity space in the local parts of the Milky Way disk, although it is mainly the \textit{over-densities} (ridges) that have been studied for now (e.g. in \cite{Antoja+18} and \cite{Kawata+18}). Their properties could be indicative on their respective origins, thus shedding light on our Galaxy structure and its evolution. We focus on this application in the remaining of the paper to illustrate our methods.
Finally, it is also important to understand non-physical gaps in our data, as caused by instrumental imperfections or errors in analysis. While these gaps will not immediately reveal new truths about our universe, identifying them in an automated way will help accelerate science.

In our conception, a gap is a region of locally lower density inside a point cloud, such that as you move away from the gap (in some or most directions) the density rises.
To be more specific, a point is in a gap if there is (at least) one straight line you can draw such that the density locally rises away from the point in both directions along that line segment.
Thus a gap can be linear or planar or hyper-planar, and is not required, in our conception, to be spatially compact.
Furthermore, we do not consider that the surrounding density, outside the gap, must be homogeneous, not even approximately. Additionally, our focus is not on finding the `emptiest' gaps (i.e. with lowest density) but the (locally) deepest or steepest gaps: We aim at building methods that can detect not only low-density gaps, but also regions or subspaces of substantial density but which lie within even higher density regions.

In that sense, our definition of a gap is in contrast to the usage of the word ``void'' as it is conventionally used in, for example, the study of large-scale structure in cosmology, where void statistics have been important \citep[see e.g.][]{lavaux2010precision,hamaus2016constraints,kreisch2021gigantes}. Voids are defined to be compact (sometimes even spherical) regions of zero or near-zero galaxy density within the distribution of galaxies, which is homogeneous and isotropic on large scales. Voids are found with high completeness with tessellations or even counts of galaxies in spherical subvolumes  \citep{neyrinck2008zobov, sutter2015vide, banerjee2016simulating}). These methods are appropriate when the goal is to find lower-than-mean density regions in a homogeneous (on large or intermediate scales) distribution. That is excellent for the distribution of galaxies in 3-space, but sub-optimal when the target is an arbitrary point cloud in an arbitrary data space, where no homogeneity can be assumed, on any scale.



\revision{Geometric and Topological Data Analysis (TDA) have proposed tools to explore the properties of structure in data space, notably by relying on derivatives and second derivatives of a density estimate. }We only briefly overview here some major topics of interest in TDA, but for a more thorough overview of topological data analysis, we refer the reader to \cite{wasserman2018topological}, and to \cite{chen2017tutorial} for a special focus on Kernel Density Estimators in this context. 
Several works have explored the problem of mode-finding (i.e. finding the maximums within the data), proposing methods, for instance using the mean-shift algorithm, to find those modes  and then perform mode-clustering \citep{ cheng1995mean, comaniciu2002multivariate,li2007nonparametric, chacon2015population}. Another concept of interest in TDA is the Morse-Smale (MS) complex \citep{morse1925relations}, which is a way to partition the data space using `critical points' (maxima and minima) and the density gradient (using `ascending flows' and `descending flows'). It has been used notably to visualize multivariate density function in \cite{chen2017statistical}. \revision{Critical points can provide starting points to find local under-densities in the distribution (see the Appendix for more details). However, while the MS complex could be used, in principle, to define gaps (e.g. looking at the gradient flows from local minima to nearby saddle points), it will become computationally challenging as the number of dimension increases, to cover the various dimension in which gaps can lie in, or will be restricted to 1-dimensional gaps.}
\\TDA is also often associated to persistent homology, which studies how topological features change as a function of scale. \revision{While persistence diagrams provide information on the topological features of a set of points (e.g. number of loops or connected components), it does not provide "localized" information (e.g. where these features are).} 
The problem of interest that is the closest to our work is the problem of ridge detection. It has been explored for instance in \cite{genovese2014nonparametric}, whose approach is very adjacent to ours: Their goal is to extend on mode-finding and uncover hidden structures, in the form of over-densities, in the data, using projections of the Hessian. \revision{However, their method focuses on over-densities, and relies on optimization to find the ridges (i.e. it does not provide a ``score'' or a metric that can be computed at any given point). Their approach will not be adaptable to find under-densities of any shape (e.g. not 1-dimensional), and can become computationally expensive compared to ours.}



Another approach related to our problem is a clustering method presented in \cite{zhang2007neighbor}: in order to identify the clusters in a set of points, the authors propose to find the valleys and use those to separate the groups. To do so, they rely on the normalized density derivative and an approximation of the local convexity of the density (similar to the approaches we propose), using non-parametric density estimation based on neighbor numbers. However, the method requires to compute the pairwise distance between each point, which will rapidly get computationally challenging as the dataset grows. 

\revision{In this paper, we present a statistic that can be computed at (almost) any point in the data space and which can be used as an estimator of `gappiness' of a region. This statistic permits the highlighting of a wider variety of gaps (in terms of shapes and dimensions) in a more generic fashion, than methods relying on critical points (or mode-finding) or ``ridge-finding'' approaches (which are, additionally, not designed originally for under-densities but for over-densities). To the best of our knowledge, there are no other methods that address the problem of gap detection and characterization in this form.  Furthermore, as we propose a statistic that can be directly computed for a given point, the finding of gaps does not require optimization per se, contrary to other methods. This results in potential computational advantages. Interestingly, our statistic can be easily reversed to perform over-densities detection (thus including `ridges')}. 

\revision{
This paper is organized as follow: Section~\ref{sec:data} presents the dataset used as an example use-case throughout this paper and motivates our gap-characterization goal in this context. We define our statistic Section~\ref{sec:method} and showcase how it performs on real data. Section~\ref{sec:discu} provides a discussion of the results we observe on this application and the advantages and limitations of our current method.

Additionally, we provide in the Appendix additional considerations and implementation notes: we propose a specific twice-differentiable density estimator with finite support (which is used throughout this paper) that can alleviate some computational issues compared to e.g. a classical Gaussian Kernel. We also comment on other possibles ways to identify and trace gaps, using notably critical points, and we provide possible methodologies to do so. These can provide different properties than the statistic presented in the main paper that can be relevant to specific usecases.} 

A Python implementation of the methods and the density estimator is available at \url{https://github.com/contardog/FindTheGap}.

\section{Example data}
\label{sec:data}

We test and demonstrate the methods presented here on Galactic velocities and positions data, where the identification and characterization of gaps are of crucial importance. We provide more details on the different datasets we build in the remaining of this Section.

The second data release \citep[DR2;][]{DR2} from the European Space Agency (ESA)’s \textsl{Gaia} mission \citep{GaiaMission} revolutionized our view of the Milky Way by providing position on the sky, parallaxes and proper motions for over a billion stars across a large portion of the Galaxy. It also provided radial velocities for around 7 million stars, mostly within a few kpc of the Solar neighborhood. This 6-D phase space sample revealed numerous disequilibrium features in the positions and kinematics of stars in the Solar neighborhood and beyond.

Such disequilibria features manifest as ridges and gaps in various dimensions. For example, \cite{Antoja+18} found a striking spiral pattern in the distribution of vertical position, $z$, vs vertical motion $v_z$, which shows that the Milky Way is still phase mixing after some vertical perturbation, e.g. the passage of a satellite such as the Sagittarius dwarf galaxy \citep[e.g.][]{Antoja+18} or the buckling of the Galactic bar \citep{Khoperskov+19a}. \cite{Antoja+18} and \cite{Kawata+18} also found ridges in the Galactocentric rotation velocity $v_{\phi}$ as a function of Galactic radius $R$, which can be signatures of the same satellite passage \citep[e.g.][]{LMJG19,Khanna+19}, Galactic spiral arms \citep[e.g.][]{Hunt+18} or resonances from the Galactic bar \citep[e.g.][]{Fragkoudi+19}, or most likely a complex combination of all three.  

These ridges, and gaps, are an extension of the long known structure in the local $v_{\mathrm{R}}-v_{\phi}$ kinematics across the observable disc. However, the change in the location of the ridges or gaps in kinematic space as a function of position in the Galaxy can shed light on their origin. For example, a gap with a resonant origin will move with a rate dependent on the order of the resonance in Galactic azimuth, $\phi$. Developing methods that can not only find gaps in phase space, but also quantify their rate of change in higher dimensional space will allow us to determine the origin of specific kinematic substructure, which in turn informs us on the structure and evolutionary history of our Galaxy. In this work we concentrate on the development and showcasing of the gap finding algorithm at work in the \textsl{Gaia} data, and defer the scientific interpretation of the substructure to future work.

\begin{figure}
    \gridline{ \fig{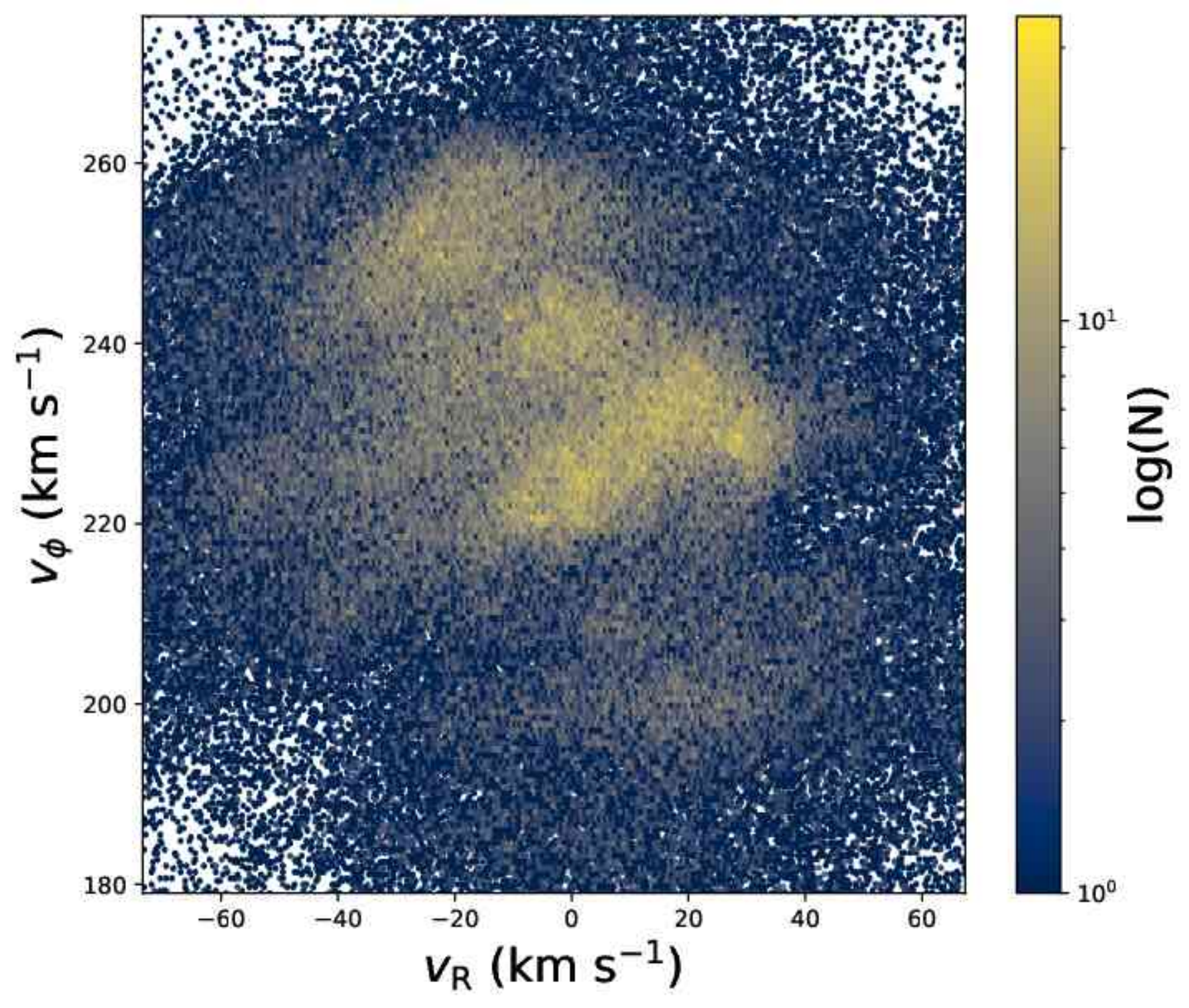}{.5\textwidth}{(a)}
     \fig{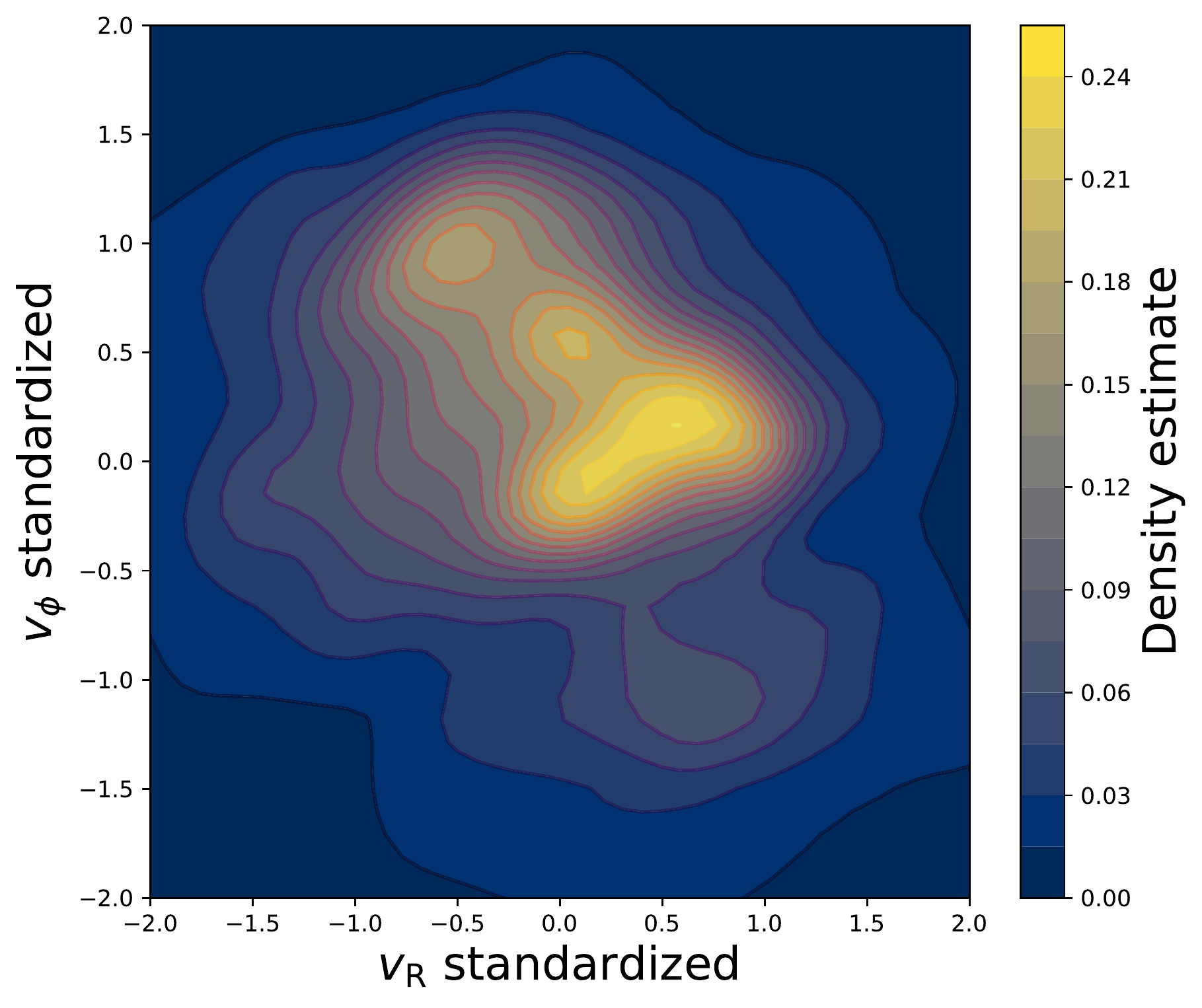}{.5\textwidth}{(b)}
    } 
    \caption{\textit{Left (a)}: 2-dimensional dataset of velocities $v_\phi$ and $v_{\mathrm{R}}$ visualized as an hexagonal binning plot.
     \textit{Right (b)}:Density estimated with our Quadratic Kernel Density Estimator \revision{(See implementation notes in Appendix)}, with a bandwidth $\Delta = 0.15$. \revision{Data is rescaled using standard normalization beforehand.} \label{fig:hex_dens}}
\end{figure}

For our sample, we use the recent intermediate data release \citep[eDR3;][]{edr3} which brings updated astrometry, but no new radial velocity measurements, which are taken from DR2. We perform the photometric quality cuts that were suggested in \cite{SME19} for DR2, namely we select stars with a color of $G_{\mathrm{BP}}-G_{\mathrm{RP}}<1.5$, a magnitude of $G<14.5$, a fractional parallax error of $\pi/\sigma_{\pi}>5$, a parallax uncertainty cut of $\sigma_{\pi}<0.1$, a BP-RP excess flux factor of $1.172<E_{\mathrm{BPRP}}<1.3$, and with more than 5 visibility periods used, which may be overkill for eDR3. We derive distances naively as $d=1/\pi$. Such an approximation is only valid for highly accurate parallaxes, yet the purpose of this work is to detect gaps, not make rigorous measurements. We use \sc{galpy }\rm \cite{B15} to convert from the \textsl{Gaia} frame $(\alpha,\delta,\pi,\mu_{\alpha},\mu_{\delta},v_{\mathrm{R}})$ to cylindrical coordinates $(R,\phi,z,v_{\mathrm{R}},v_{\phi},v_z)$ assuming a distance to the Galactic centre of $R_0=8.178$ kpc \citep{Gravity+19}, and the Sun's height about the disc plane as 20.8 pc \citep{BB19}. We calculate the vertical and azimuthal Solar motion by combining $R_0$ with the proper motion measurement of Sgr A* of $(\mu_l,\mu_b)=(-6.411\pm0.008,-0.219\pm0.007)$ \citep{RB20}. Thus we have $v_{\odot}=248.5$ km s$^{-1}$ and $w_{\odot}=8.5$ km s$^{-1}$. Finally, we perform an additional cut on the velocities, selecting stars with $-125 < v_{\mathrm{R}} < 125$ and $100 < v_\phi < 300$. 

\begin{figure}
        \centering
        \includegraphics[width=\textwidth]{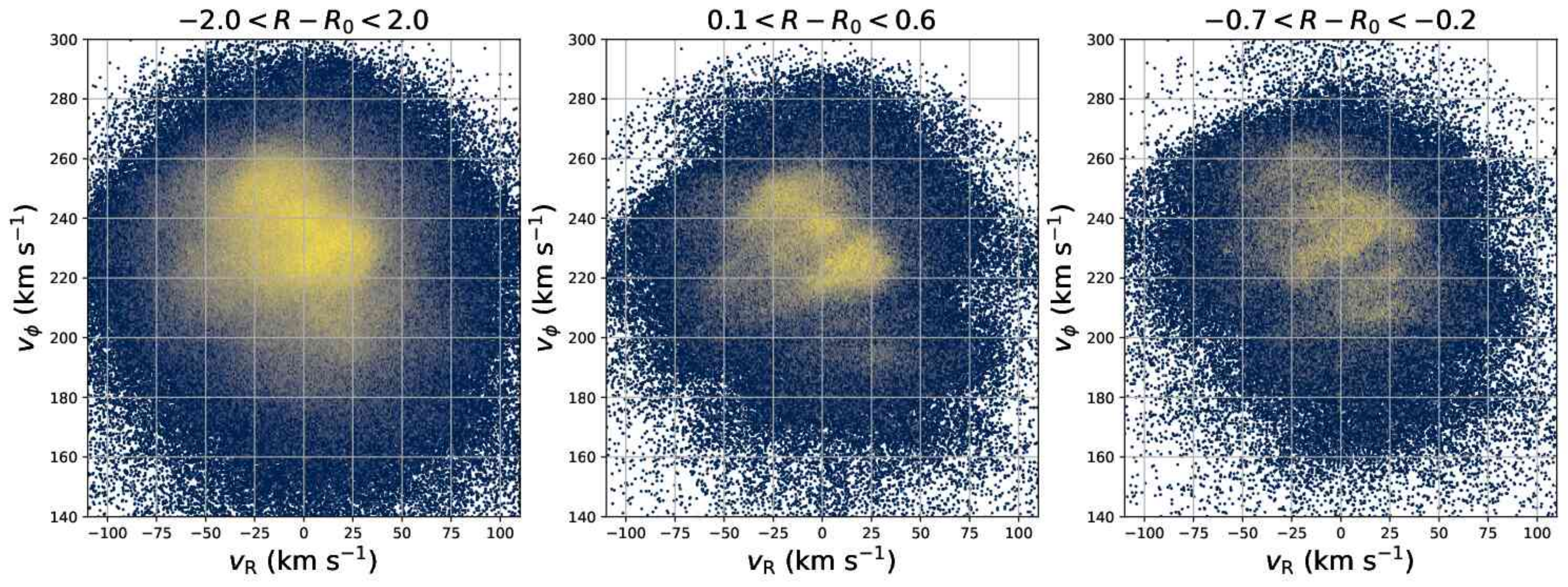}
        \caption{Projection of the 3-dimensional dataset $D_R$ (features $v_\phi, v_{\mathrm{R}}, R-R_0$) in velocity space, with all data (left panel), stars within $ 0.1 < R - R_0 < 0.6$ (middle panel), and stars within $ -0.7 < R - R_0 < -0.2$ (right panel). Gaps in the distribution are most visible in the sub-sampled data (middle and right panel) and are visibly different for the two cuts, indicating a dependency with $R - R_0$.
        \label{fig:data_3d_R_slicehex}}
   
\end{figure}

From these data, we create the following datasets: 
\begin{itemize}
    \item A 2-dimensional dataset $D_v$, using $v_{\mathrm{R}}$ and $v_{\phi}$, with an additional cut selecting stars within a distance of 200 pc from the Sun, resulting in a dataset of $\sim$217,000 stars. Visualization of this dataset is provided in Figure \ref{fig:hex_dens}a.
    \item Two 3-dimensional datasets, adding $R$ and $\phi$ as a third-dimension respectively:
    \begin{itemize}
        \item $D_R$ using $v_{\mathrm{R}}, v_\phi$ and $R-R_0$, with an additional cut selecting stars within 2 kpc in $R - R_0$ and 300 pc in $\phi$, resulting in a dataset of $\sim$ 910,000 stars. 
        Figure \ref{fig:data_3d_R_slicehex} shows the entire dataset $D_R$ in the velocity space as well as two subset of the data in sub-windows in $R - R_0$. The gaps in those subsets are more visible, and one can see that their locations and shape (e.g. width) change from one subset to the other. 
        \item $D_\phi$  using $v_{\mathrm{R}}, v_\phi$ and $\phi$, with an additional cut selecting stars within 200 pc in $R - R_0$ and within $15^{\circ}$ in $\phi$, resulting in a dataset of $\sim$ 752,000 stars.
    \end{itemize}
    \item A 4-dimensional dataset $D_{R,\phi}$ including both velocities $v_{\mathrm{R}}, v_\phi$ and both positions $R-R_0, \phi$. We keep stars within 2 kpc in $R$ and within $15^{\circ}$ in $\phi$, resulting in a dataset of $\sim$ 1,670,000 stars. 
\end{itemize}
\revision{We re-scaled all datasets before computing density estimations and our methods throughout the paper, using standard normalization (other scaling methods could be used), in order to ensure coherent scale across dimensions. The methods presented here assume that all dimensions in the data are properly rescaled for the task at hand and/or according to underlying assumptions regarding the data. This might play a crucial role to find structures in the density distribution as our kernel is symmetric in all dimensions.}

\section{Method}
\label{sec:method}

\begin{figure*}
\gridline{\fig{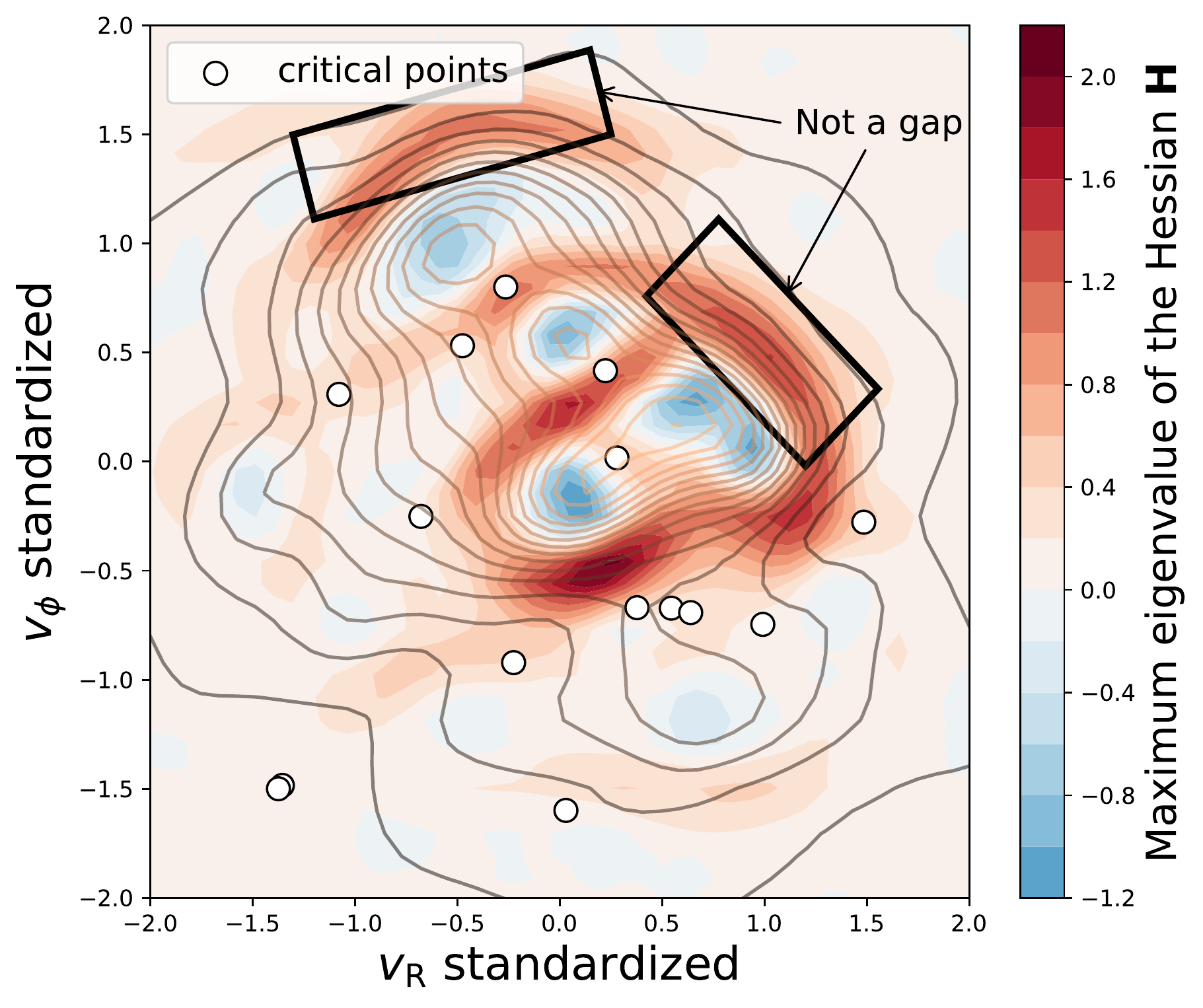}{0.5\textwidth}{(a)}
          \fig{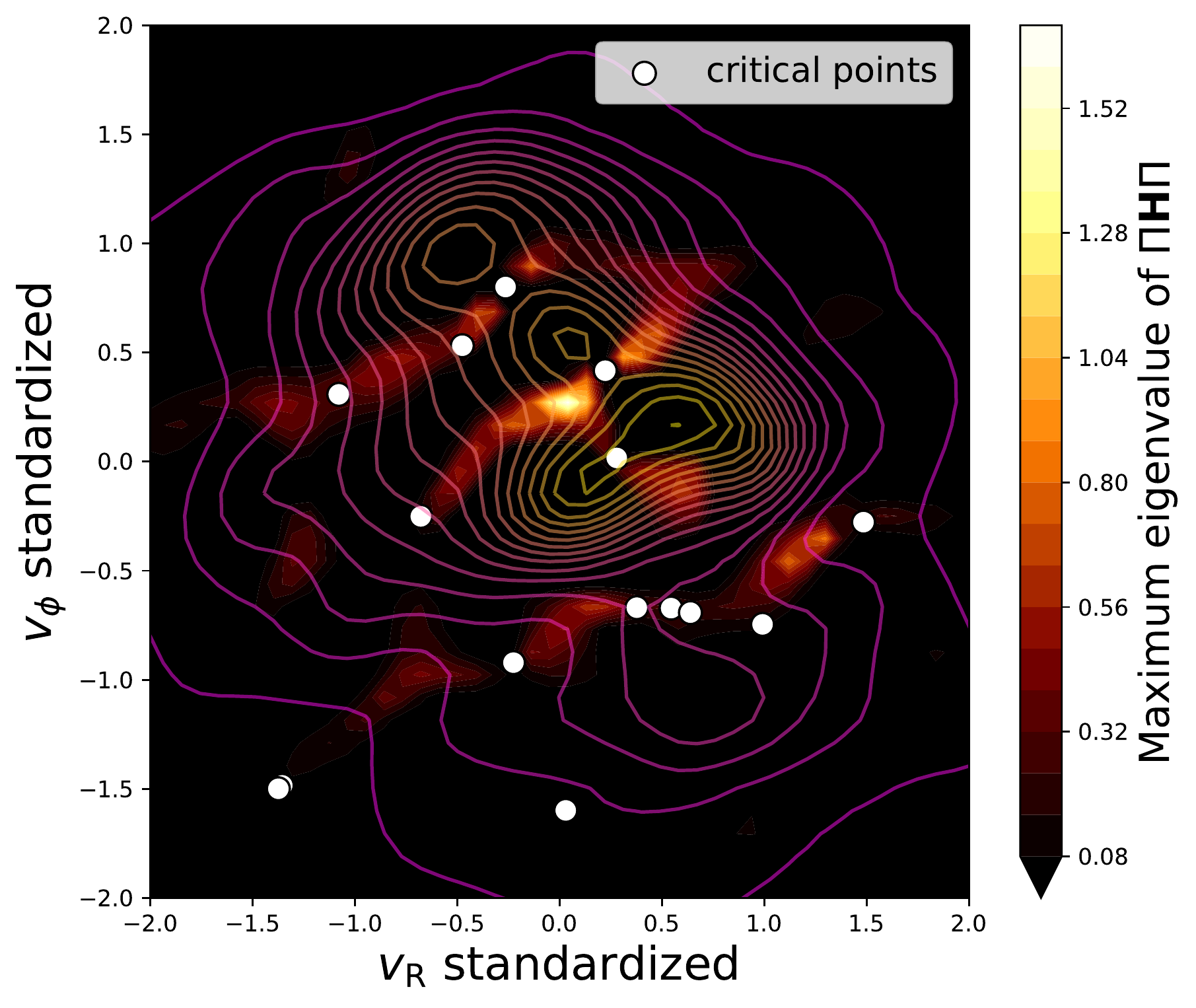}{0.5\textwidth}{(b)}
          }
\caption{\textit{Left (a):} \textbf{Map of the maximum eigenvalue of the second derivative (Hessian) of the density} estimated with a bandwidth $\Delta = 0.15$ on a grid in the data space of $D_v$, with density estimate shown as contour lines. We highlight a region with high values for $\lambda \mathbf{H}$ that is however not a gap in the density distribution according to our definition, as it is not between two regions with higher density. White circles show the 15 critical points with the highest Hessian's maximum eigenvalues (see Appendix for details).
     \textit{Right (b):} \textbf{Map of the maximum eigenvalue of \pihpi} for a bandwidth $\Delta = 0.15$ on a grid in the data space of $D_v$, with density estimate shown as contour lines. This criterion, compared to the maximum eigenvalue of the Hessian alone, efficiently removes areas with steep curvatures in the density that are however not gaps. \label{fig:lambdaH_PiHPi}}
\end{figure*}


\revision{
This Section details the statistic we propose to estimate the ``gappiness'' of a point in the data space. 

As a reminder, our goal is to provide a measure or statistic that can be interpreted as ``how much a point can be considered to be lying within a gap'', for any point in the space. Our definition of a gap is a region of \textit{locally} lower density, such that there is (at least) one straight line along which density rises as we go away from the region in either direction along that line. 

This definition is more generic than for instance relying on critical points, which are points where the gradient of the density is 0 (and could be minima, maxima or saddle points) (see Appendix for more notes on that aspect). While a gap will have (at least) one critical point, its structure or surface can extend beyond that critical point. Our definition, without additional constraints, potentially extends a ``gap'' arbitrarily, as long as there is density rising on both side.

Let us consider a dataset (or point-cloud) $D$ of $n$ points $x_i \in \mathbb{R}^d$, with boundaries (limits in $\mathbb{R}^d$) $S$. 
The approach we propose relies on a density estimator that can be of any nature, as long as it is twice-differentiable. In the remaining of the paper, we consider a kernel-based density estimator  (see Appendix for the specific implementation used in our experiments), but the proposed statistic is agnostic to the actual estimator. We denote the density estimator $p$. As we consider kernel-based estimator, $p$ is associated with a bandwidth that we denote $\Delta$. The estimator $p_{\Delta}$ is fitted on all the points within $D$. For any given point $x \in \mathbb{R}^d$, it is possible to compute the density estimate of that point and the gradient and second derivative (Hessian) of that density estimate. We denote the gradient vector of the density estimate \gradnotation, and the Hessian matrix $\mathbf{H}$, dropping the notation of the point of interest $x$ for simplicity.


Following our definition of a gap, it makes sense to look at the properties of the second derivative (Hessian) of the density, as it is an indicator of the curvature of the density field. The maximum eigenvalue of the Hessian $\mathbf{H}$ gives us an estimate of the regions of the space that have the biggest positive curvature in the density field: this is illustrated in Figure \ref{fig:lambdaH_PiHPi}(a) on our 2-dimensional dataset. Negative values corresponds to ``peaks'' or ``islands'' in the density. High positive values align with regions where the density drops sharply (at that given bandwidth). These can be ``valleys'' (matching our gap definition), or they can be ``cliffs'': edges where the density do not rise in any other direction (highlighted by black rectangles in Figure \ref{fig:lambdaH_PiHPi}(a)). These regions are not considered gaps in our definition.

Therefore, we need to examine the properties of the Hessian in a slightly different fashion, to account for this. 
} The second derivative makes most sense to examine along directions where there is no first derivative; that is, where the second derivative delivers the most important non-trivial term in the Taylor series.
Put another way, the second derivative confined to the subspace perpendicular to the gradient vector is the second derivative along all directions in which there is no first derivative.
Because of that, every point in the $d$-dimensional data space can be thought of as a critical point (point of zero first derivative) in the $(d-1)$-dimensional subspace locally perpendicular to the local gradient at that point.
Thus it makes sense to examine the eigenvalues of the Hessian in that local subspace to ask whether the point is a local underdensity or local overdensity in the subspace.
For our gappiness statistic, therefore, we propose to use the eigenvalues of the projection of the Hessian into the orthogonal subspace of the density gradient.

In that sense, our statistic relies on similar motivations than the ridge-finding optimization method presented by \cite{genovese2014nonparametric} (low-dimensional over-densities in data).
For $1 \leq d' < d$ (where $d$ is the number of dimension of the dataset), a $d'$-dimensional gap is characterized by a Hessian with $d - d'$ large, positive values, and where the projection of the gradient on that subspace is 0.

We denote this projected second derivative tensor \pihpi, where $\mathbf{H}$ is the Hessian matrix, and where $\Pi$ is a projection operator that projects into an orthogonal subspace, orthogonal to the gradient vector \gradnotation.
If \gradnotation is seen as a column vector, then
\begin{equation}
    \Pi \equiv \mathbb{I} - \frac{\nabla p\,{\nabla p}^\top}{ {\nabla p}^\top {\nabla p} } ~,
\end{equation}
where $\mathbb{I}$ is the identity.
This measure \pihpi\ can be computed at any point in the data-space.  However, it will be undefined precisely at the true critical points (where the gradient vanishes), and indefinite for any truly empty gap (zero density at a given bandwidth).

We also note that other statistics, over \pihpi or its eigenvalues, could be interesting to explore, such as e.g. the trace of \pihpi, which would provide different properties, especially at higher dimensions. Using the minimum eigenvalues of \pihpi, on the other hand, will provide a ``ridge'' (or local over-densities) statistic.

Figure \ref{fig:lambdaH_PiHPi}b illustrates the map of the maximum eigenvalues of \pihpi for all the points on a grid for the 2D dataset $D_v$, with the density estimate shown as contour lines, as well as the 15 critical points with the highest maximum eigenvalue of the Hessian (see Appendix). We can see that this measure efficiently eliminates the irrelevant regions compared to using the maximum eigenvalue of the Hessian only, and highlights the gap-regions relevant to our definition. This visualization also shows that the regions with high values aligns with concave areas in the contour lines of the density, as expected given our target definition of a gap. We can see however that some of the `wider' gap regions (e.g. near 0.0 in $v_{\mathrm{R}}$ and $-1.0$ in $v_\phi$) are not entirely covered and evaporate. This is related to the relationship between our gap-statistics and the bandwidth $\Delta$ used by the density estimator. 

Indeed, the choice of bandwidth will be critical to the properties of the gaps detected, as it will play a role e.g. in the size of the gaps that can be detected. Deciding on a single bandwidth to do the analysis might be non-trivial, and might leave out some important gaps. \revision{Small bandwidth might allow to find smaller gaps but they might be more sensitive to noise as well and will tend to under-smooth. This can make `spurious' gaps might appear. On the other hand, large bandwidth will tend to over-smooth and some gaps will disappear. There is also a relationship with the number of datapoints in the dataset. The problem of bandwidth selection for kernel-based density estimators has been actively studied and several methods have been proposed, such as Scott's rule of thumb \citep{scott2015multivariate}, Silverman's rule of thumb \citep{silverman1998density}, or Sheather and Jones method \citep{sheather1991reliable}, among others. However, the problem remains unsolved in general.} Therefore, ensuring the stability of our statistic evaluating the ``gappiness'' of a region / point in the dataspace is crucial. We propose to explore summary statistics of the maximum eigenvalues of \pihpi when ran across different bandwidths, and using bootstrapping.

\begin{figure}
    \centering
    \includegraphics[width=\textwidth]{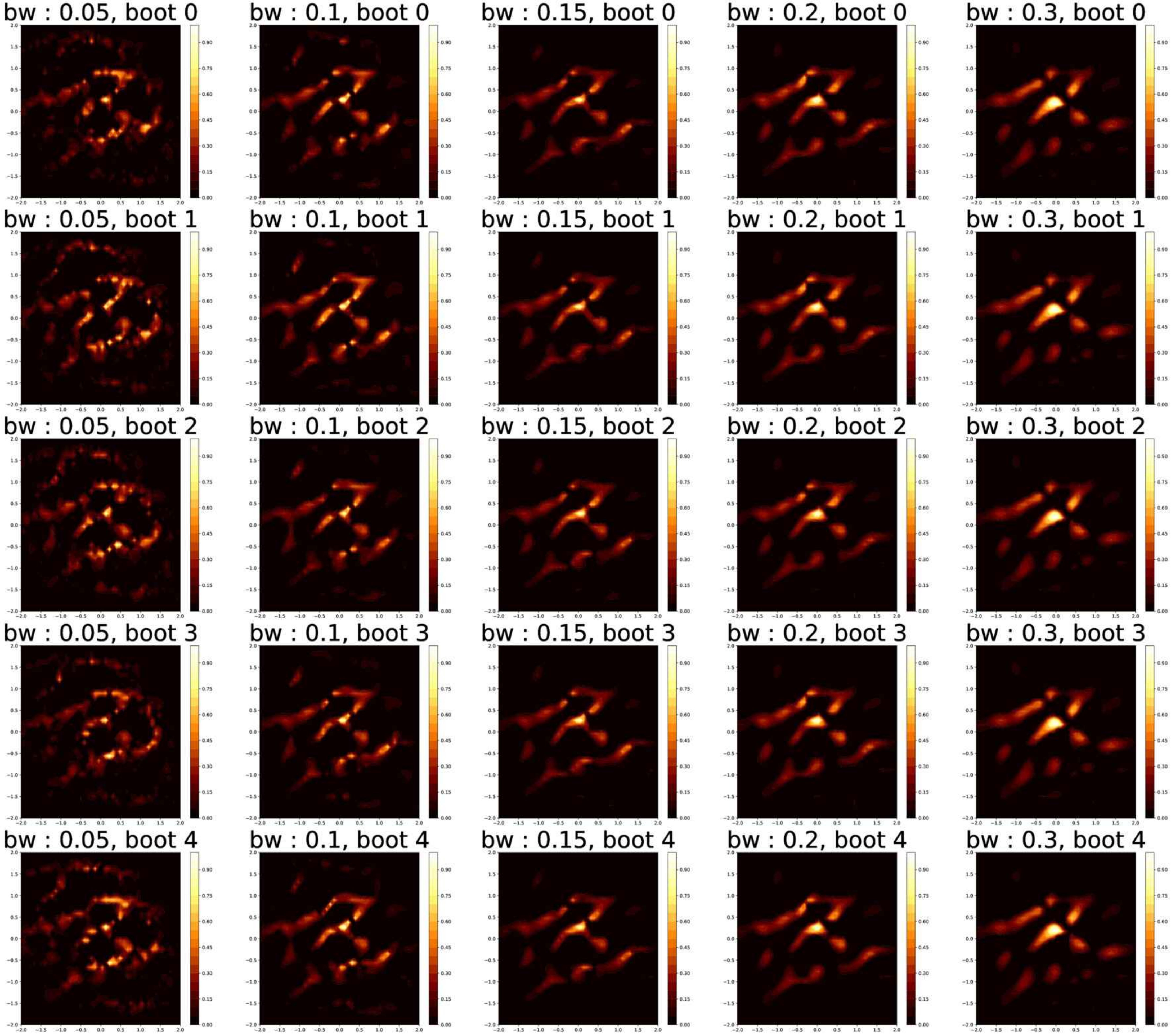}
    \caption{Map of the maximum eigenvalue of \pihpi for different bandwidths $\Delta$ (columns) and different bootstrapping sampling (rows). Smaller bandwidths can grasp finer and smaller gaps structures (e.g., bandwidth of 0.1), however extremely low bandwidth becomes susceptible to find spurious `gaps', which will not be stable or consistant across bootstraps (e.g., plots in the first column). At larger bandwidth, detected gaps become more stable across bootstraps, and the criterion can detect larger and wider structures. }
    \label{fig:lambda_PiHPi_gridbwboot}
\end{figure}

Figure \ref{fig:lambda_PiHPi_gridbwboot} shows the maximum eigenvalue of the \pihpi criterion for different bandwidth (columns) and different sampling (rows). It is apparent there that at the smallest bandwidth, the gap-estimation is dominated by noise, hence unstable per point. At larger bandwidth, some gaps disappear entirely, but the wider `valleys' get a better and more stable coverage. 

\begin{figure}
    \centering
    \includegraphics[width=.5\textwidth]{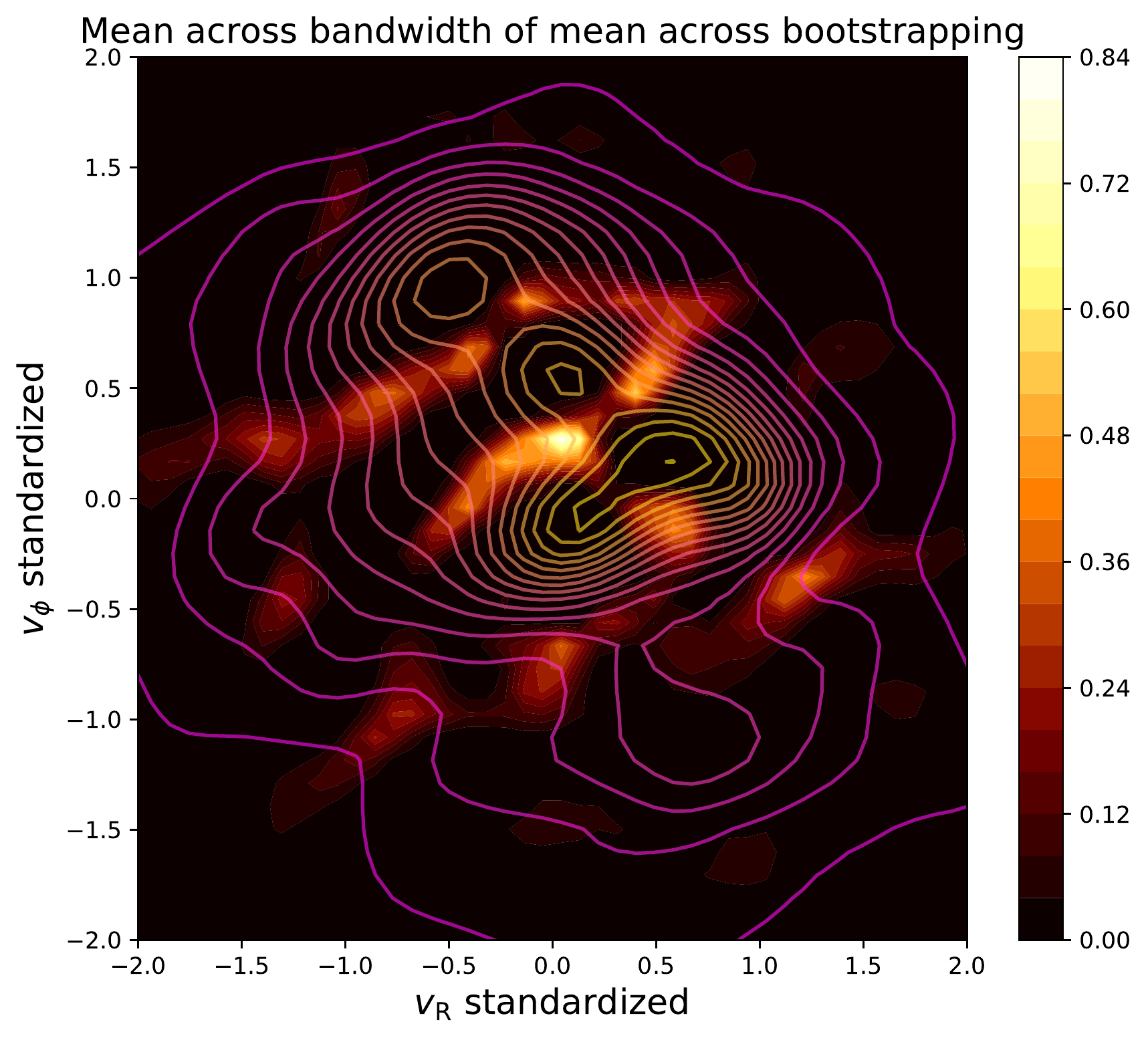}
    \caption{Map of the mean of the maximum eigenvalue of \pihpi (rescaled between 0 and 1 for each run) across runs with different bandwidths and bootstrapping (see Figure \ref{fig:lambda_PiHPi_gridbwboot}). Taking summary statistics across bandwidths and bootstrapping allows to highlight gaps of different sizes while keeping only the stable gap regions.}
    \label{fig:lambda_PiHPi_summary}
\end{figure}

\revision{By taking summary statistics across the different runs, it is possible to alleviate the instability of some gaps and to combine the stable gaps of different widths. First, we rescale each gap-measurement map (maximum eigenvalue of \pihpi) for each individual run so that its values lie between 0 and 1. Then, we take the mean across the different bootstrap run, in order to remove the `unstable' gaps at smaller bandwidth. Finally, computing the mean of the mean-maps across bandwidth allows to keep the stable gaps of different width. We show in Figure \ref{fig:lambda_PiHPi_summary} the final summary map averaging all the rescaled maps shown in Figure \ref{fig:lambda_PiHPi_gridbwboot}. Compared to Figure \ref{fig:lambdaH_PiHPi}b, using a summary across bandwidths allows to better recover the wider gaps. However, we note that in more general cases, users want to be careful in merging the statistic across bandwidths, as, in some instances, a bigger bandwidth might merge two smaller gaps together.}

We also note that, although we explicitly perform bootstraps in Figures \ref{fig:lambda_PiHPi_gridbwboot} and \ref{fig:lambda_PiHPi_summary}, it is probable that explicit bootstrap is not required here to assess the shot noise (sample variance): the density estimate (and its first and second derivatives) are weighted sums over neighbors. Since we explicitly have a sum over weights to construct the density and its derivatives, we could look at the statistics of those weights to determine the shot noise variance in the density estimate. That would deliver further speed improvements to any pipeline making use of these tools.


\begin{figure} 
    \gridline{
        \fig{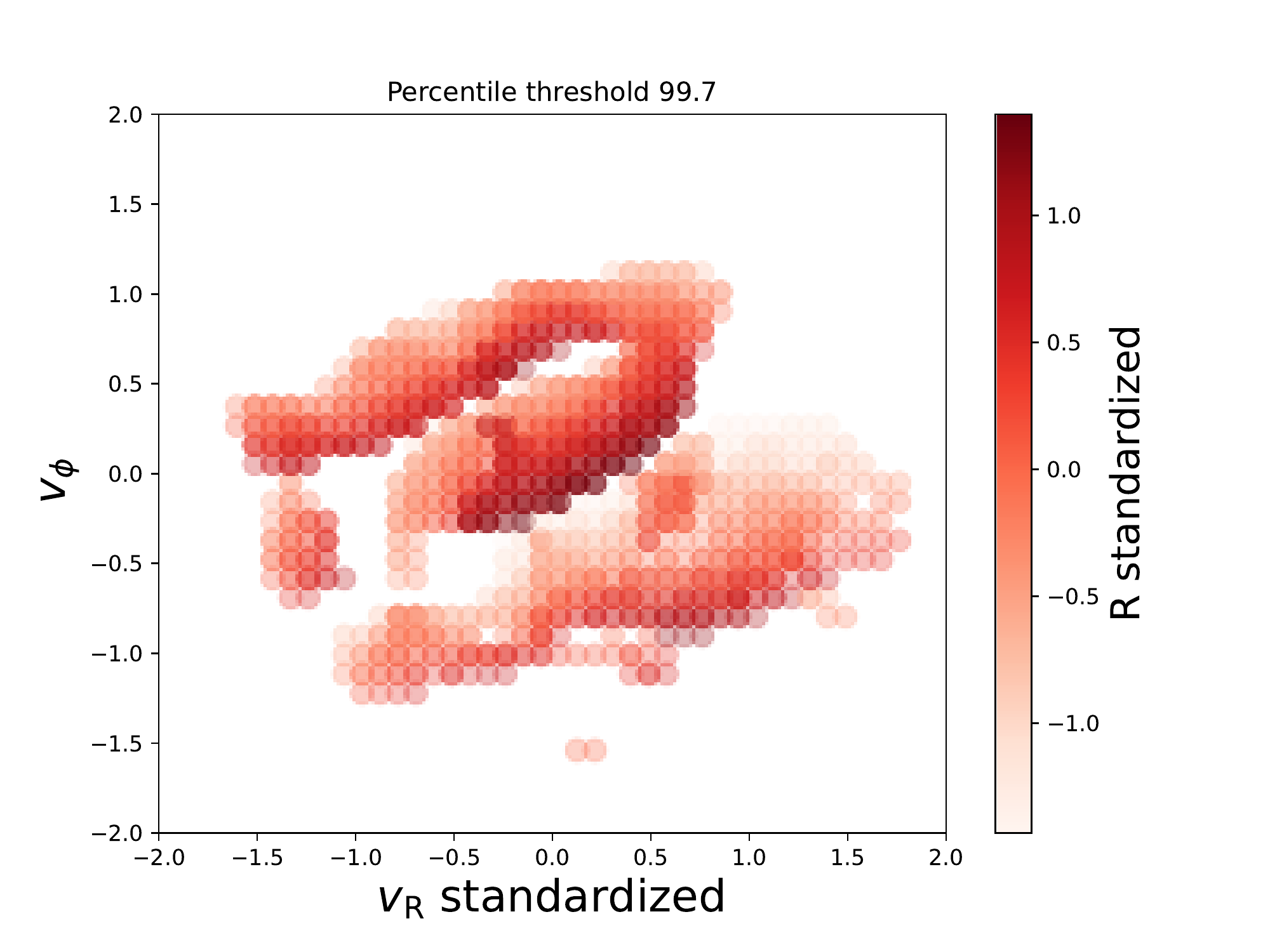}{.5\textwidth}{(a)}
        \fig{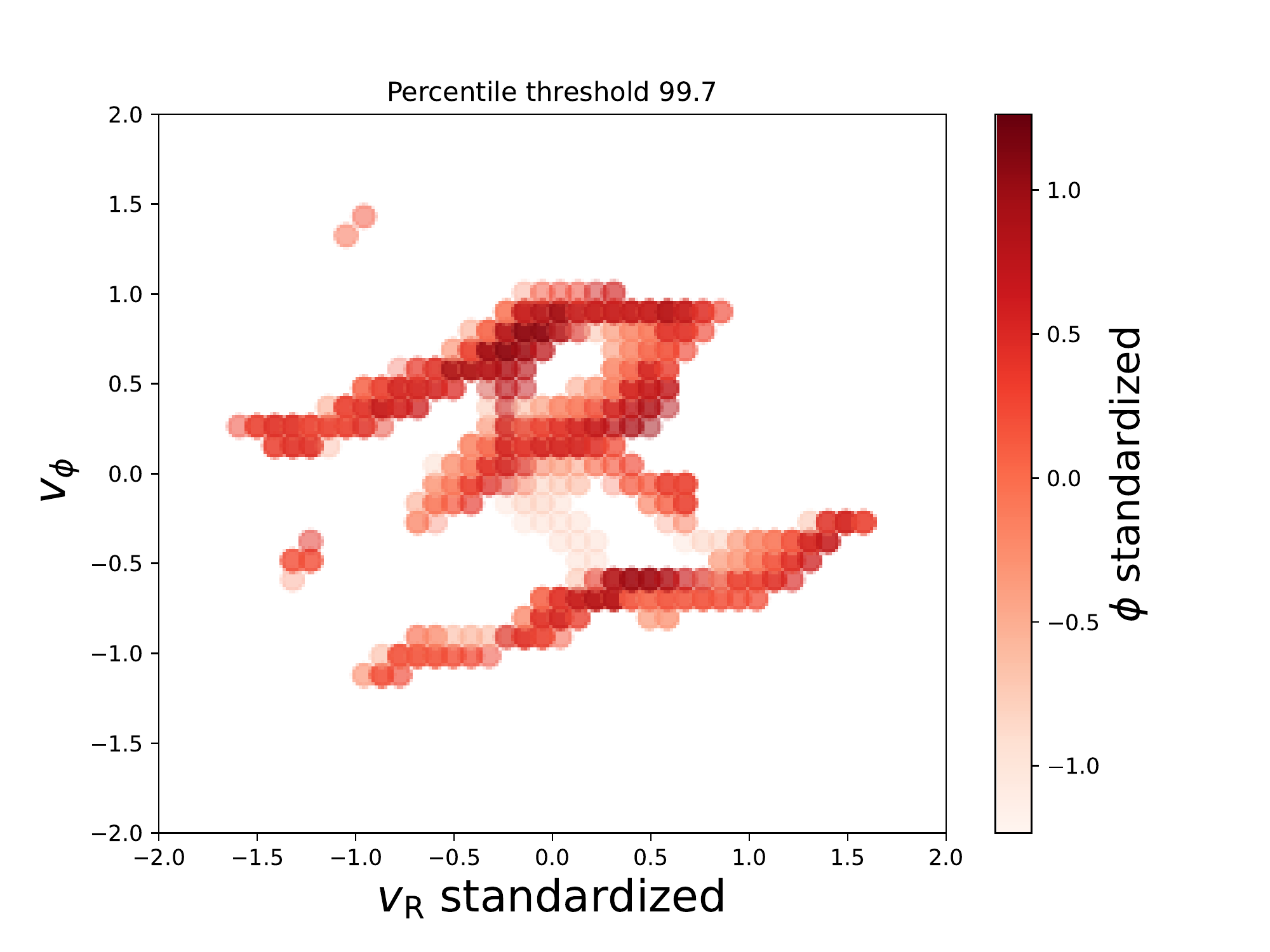}{.5\textwidth}{(b)}
    }
    \caption{\revision{Regions in the 3-D datasets $D_R$ (left (a)) and $D_\phi$ with a maximum eigenvalu of \pihpi above the 99.7th percentile on a $80 \times 80 \times 80$ grid, using a Quadratic Kernel Density estimator with a bandwdith of $0.15$. Selected regions are projected on the 2-D velocity space and colored by their third dimension $R$ and $\phi$ respectively.}  \label{fig:pihpi_3d}}
\end{figure}

\revision{Similarly, our ``gap'' statistic can be computed on our 3-dimensional datasets by gridding them, and using the statistic as a selection threshold to highlight which region are gaps in the distribution. Figure \ref{fig:pihpi_3d} shows the regions in $D_R$ and $D_\phi$ respectively with a maximum eigenvalue of \pihpi above the 99.7th percentile, on a $80 \times 80 \times 80$ grid, with our Quadratic kernel density estimator, with a bandwidth of $0.15$. The selected regions are visualized in two dimensions $v_R$ and $v_\phi$, colored by their third dimension ($R$ and $\phi$ respectively). This clearly shows several gaps, that move strongly as a function of R, and more subtly in $\phi$. }

 \revision{In many use-cases and applications, it might be interesting to focus instead on which \textit{data-points} lie within gaps. This might prove useful from a scientific or data-exploration point-of-view (studying the distribution and properties of the data lying in underdensities). It also provides an indirect way to trace the gap regions in a possibly less computationally expensive way than a fine-enough gridding would require, especially on higher-dimensional datasets. However, this ``trick'' will obviously fall short in the case of trully empty gaps.} 

Our gap statistic can be applied similarly to the data-points within the dataset directly. Depending on use-cases, one can decide to keep all data-points above a certain value of the maximum eigenvalue of \pihpi depending on its distribution \revision{(we note that this statistic is not a normalized statistics) }, to keep the data-points above some percentile, or other criterion for cut. For sake of illustration and validation, Figure 



\begin{figure}
    \gridline{
    \fig{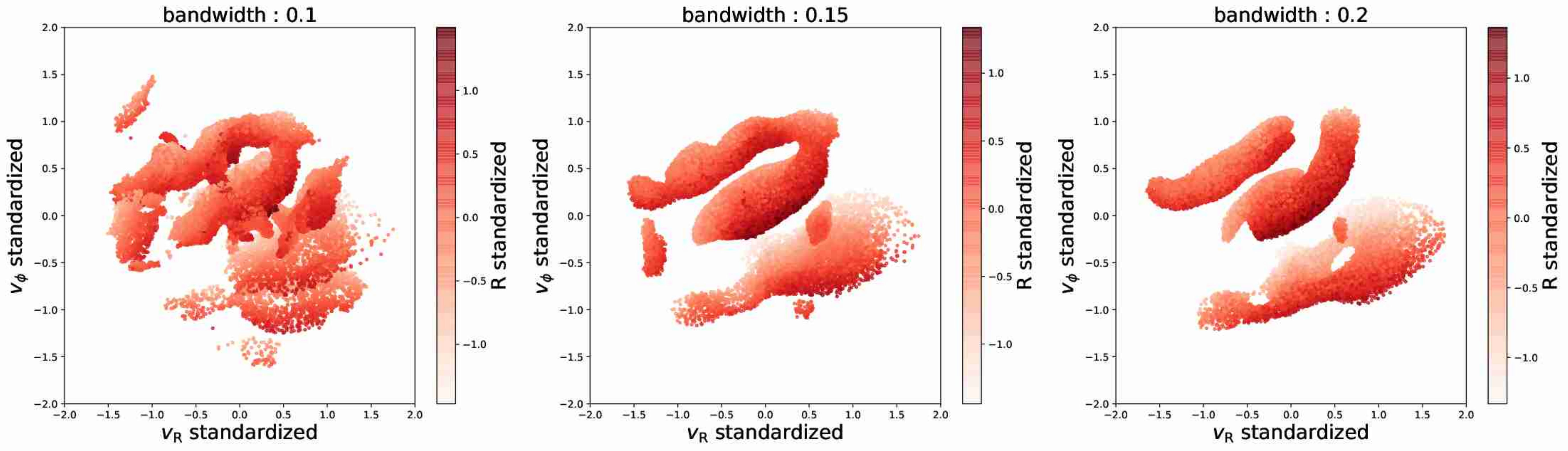}{.9\textwidth}{(a)}
    }
    \gridline{
    \fig{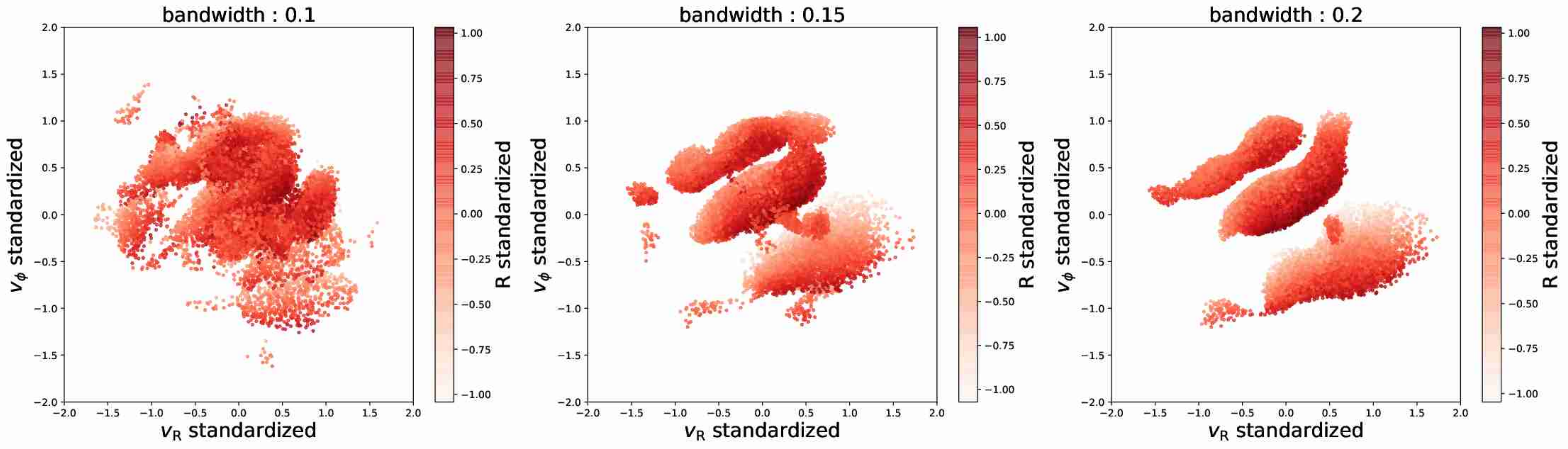}{.9\textwidth}{(b)}
    }
    \caption{Data-points selected in the 3-D dataset $D_R$ (top (a)) and the 4-D dataset $D_{R,\phi}$ (bottom (b)), such that their maximum eigenvalue of \pihpi is above the 95th percentile (top) and the 98th percentile (bottom). Points are colored by R.
    \label{fig:3-4D_R_ano}
    } 
   
\end{figure}
\begin{figure}
    \gridline{
    \fig{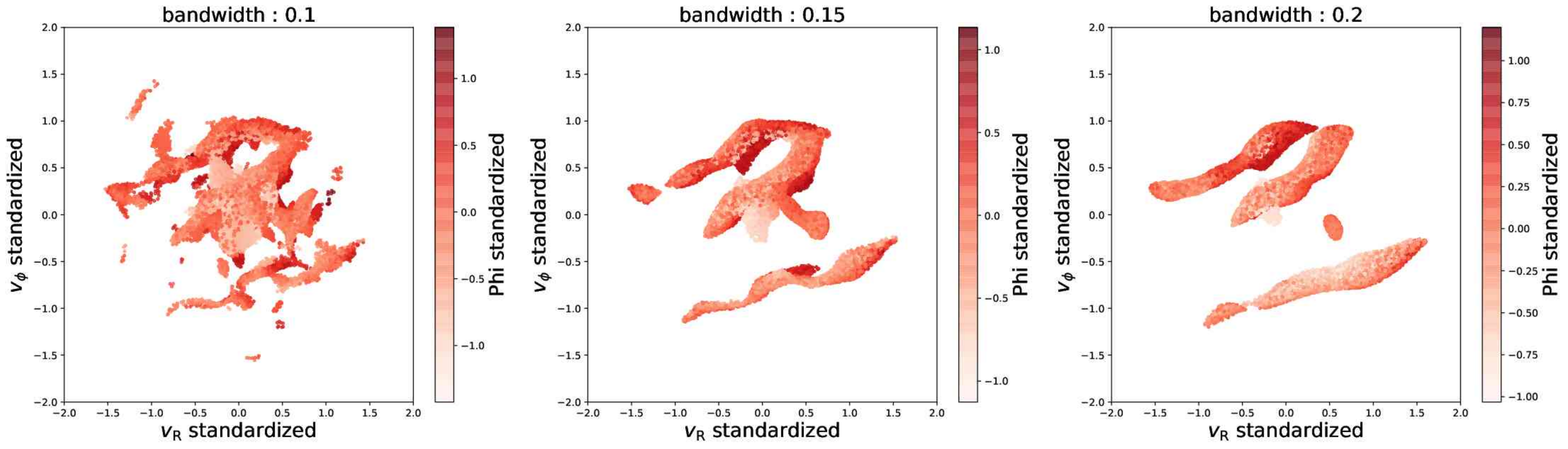}{.9\textwidth}{(a)}
    }
    \gridline{
    \fig{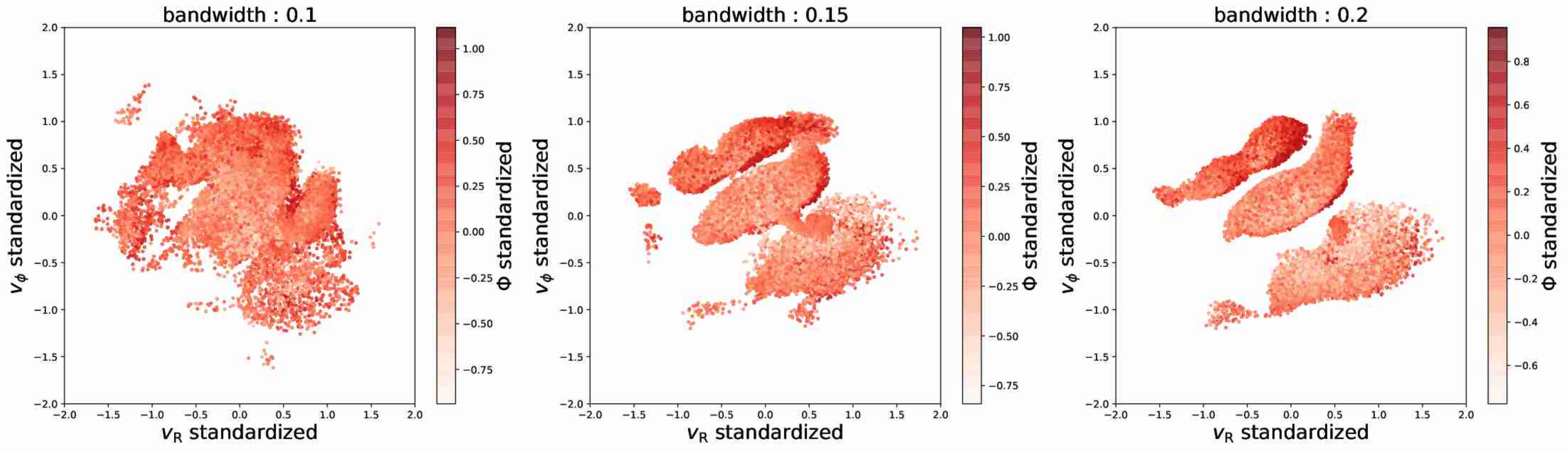}{.9\textwidth}{(b)}
    }
    \caption{Data-points selected in the 3-D dataset $D_\phi$ (top (a)) and the 4-D dataset $D_{R,\phi}$ (bottom (b)), such that their maximum eigenvalue of \pihpi is above the 95th percentile (top) and the 98th percentile (bottom). Points are colored by $\phi$.
    \label{fig:3-4D_phi_ano}}
\end{figure}

We apply this selection process on our 3-dimensional datasets $D_R$ and $D_\phi$, integrating the positions in $R$ and $\phi$ individually. Figures \ref{fig:3-4D_R_ano}a and \ref{fig:3-4D_phi_ano}a respectively show the datapoints above the 95th percentile for the maximum eigenvalue of \pihpi, for three different bandwidths, with the points colored by their respective additional position features $R$ and $\phi$. We see very similar pattern as in Figure \ref{fig:pihpi_3d}, with gaps moving strongly as a function of $R$, and more complex relationship with $\phi$. We can apply a similar approach to our 4-dimensional dataset $D_{R,\phi}$, combining both $R$ and $\phi$ position. Figures \ref{fig:3-4D_R_ano}b and \ref{fig:3-4D_phi_ano}b shows the 98th percentile selection on our gap criterion computed with three different bandwidths (similar to the previous Figures), respectively colored by $\Phi$ and $R$ as well. We can see that the `clusters' formed by the selected points become more blurry. While we show results with the same three bandwidths for the 3-dimensional and 4-dimensional cases for sake of illustration, it is unlikely that the `best' bandwidth for the 3-D case will translate to the 4-D case.

\section{Results and discussion}
\label{sec:discu}

\revision{We present in this paper a statistic that can be used to highlight gaps, or local under-densities, in data distribution. We showcase the ability of our statistic to retrieve the observed gaps in the velocity distribution of nearby stars in the Milky Way, and its potential to gain insights on the properties of those gaps. Our method relies on the use of a twice-differentiable density estimator. Such methods might themselves depend on a choice of bandwidth, which in turn impacts the possible gaps found by those methods. We illustrate that it is possible to combine different bandwidths sensibly to detect gaps of various widths, combined to a methodological way to ensure robustness of the gaps detected, using bootstrapping. When applied to our 3 and 4-dimensional datasets, we confirmed our expectations that the gaps in the kinematic space have dependencies with $R$ and $\phi$ in terms of position and slope evolution. We however defer further investigation on the underlying physics driving those gaps to future works. }


Several paths for future investigation still remain. First and foremost, it remains non-trivial to go to much higher dimensions: one limitation will come from the natural limits of (kernel) density estimators to provide confident density estimate on higher dimensional datasets. Besides this crucial problem, visualizing and extracting the characteristics of the gaps detected by our methods, when in higher-dimensions, might become non-trivial as well. 

Additionally, we presented here an analysis and tools using specific choices of `gap-criterion', that approximate well our gap definition. But it would be interesting to investigate other possible criterions, and statistics that could be computed on the gaps and under-densities. For instance, characterizing the `depth' of a gap instead might be relevant. However, such a measurement will be non-trivial to define and to compute efficiently. Another limitation of our methods might lie in the `summary' across bandwidths: in doing so, small gaps (in size) might merge and become difficult to distinguish. Exploring protocols to better handle those cases (e.g. in terms of bandwidth choice) will be crucial for some applications.

Conversely, similar investigation of `bumps' and ridges might also be of interest in many fields of applications. This problem has been more investigated in topological data analysis, applied for instance on cosmological dataset to identify and characterize filaments of the cosmic web, as in e.g. \cite{xu2019finding}. Our gap criterion as the maximum eivengalue of \pihpi can easily be reversed by taking the minimum eigenvalue instead, which will highlight the ridges and bumps in a density distribution. 

\paragraph{Acknowledgements:}
It is a pleasure to thank Dan Foreman-Mackey (Flatiron) and Soledad Villar (Johns Hopkins University), for valuable discussions.
Some of the ideas in this project were workshopped in the Astronomical Data Group Meetings at the Flatiron Institute and at the Machine Learning for Astronomy meeting at Max Planck Ringberg.

\paragraph{Software:} This work used and benefited from the following Python libraries: astropy \citep{astropy:2018}, galpy \citep{bovy2015galpy}, jupyter \citep{PER-GRA:2007}, matplotlib \citep{Hunter:2007}, numpy \citep{harris2020array}, pyTorch \citep{pytorch}, sklearn \citep{scikit-learn}, scipy \citep{scipy}.


\bibliography{biblio}

\bibliographystyle{aasjournal}

\appendix 

\section{Implementation Notes and Additionnal Considerations}

\revision{In this Appendix, we provide specific implementation notes, and additional considerations on ways to find and characterize under-densities in data distribution that might be useful to readers in specific usecases.

We first describe a specific kernel for Kernel Density Estimation that approximates a Gaussian kernel while providing finite support for faster computation. Then, we provide some comments on critical points, and how they can be used to identify and trace specific types of gaps. We propose an algorithm for approximating the critical points, and illustrate on our application how using the gradient of the density estimate, one can trace `1-D' valley gaps.}


\subsection{A fast, twice-differentiable density estimate}

\revision{The statistic presented in this paper, as well as the methods presented below in this Appendix, rely on the use of a twice-differentiable density estimator. The methods described are independent of the nature of the density estimator, so one could use whichever estimator that seem best for their applications, for instance Kernel Density Estimate, or even Normalizing Flows, as long as gradients and Hessians are available. 

However, as a practical consideration, a Gaussian kernel for instance, might become very slow for large datasets as it is not compact. On the other hand, compact kernels, leading to faster KDEs, are generally not twice-differentiable at their edges. To address this issue, we propose to create a kernel that has finite support, is twice differentiable everywhere, and which conforms to conventional ideas about bandwidth and resolution. It is the kernel used for all experiments shown in this paper.

We propose a custom Quadratic kernel, that approximates a Gaussian kernel near its center, while offering finite support and being twice differentiable everywhere, including at the edge of the support.
This density estimator is not properly normalized---it is designed to mimic a normalized Gaussian density estimator at the center of the kernel, rather than in the integral over the kernel. Nothing in our method relies on the density estimator being normalized, so this doesn't matter for our purposes, but if this density estimator is used for some integration or probability applications it would make sense to normalize it correctly instead.}


Namely, we build a kernel that is a close approximation to a Gaussian at small separations, but goes smoothly to zero, and has its slope and second derivative go smoothly to zero at what would be three-sigma in the original Gaussian.
This kernel $k(s)$ can be expressed as a quadratic polynomial:
\begin{align}
    k(s) &\equiv \left\{ \begin{array}{lll}\displaystyle
    1
    - 6\,\left(\frac{s}{3}\right)^2
    + 8\,\left(\frac{s}{3}\right)^3
    - 3\,\left(\frac{s}{3}\right)^4 & \mbox{for} & s \leq 3 \\
    0 & \mbox{for} & s > 3
    \end{array}
    \right. \\
    s &\equiv \sqrt{\frac{(x-x')^\top\,(x-x')}{\Delta^2}} \\
    K(x-x') &= \frac{1}{(2\pi\,\Delta^2)^{d/2}}\,k(s) ~,
\end{align}
where $s$ is a dimensionless separation between a data point at $d$-dimensional location $x$ and an evaluation point at $d$-dimensional location $x'$ (computed with a Euclidean metric), $\Delta$ is the bandwidth.
Implicitly $x$ and $x'$ are $d\times 1$ column vectors.
The factor of $(2\pi\,\Delta)^{d/2}$ normalizes the kernel comparably to the Gaussian it matches at small separations.
The integral of the kernel over space is not unity.

Density estimation with this kernel can be made fast, because the kernel has compact support:
In order to compute the density at a position $x$, it is only necessary to find neighboring data points out to separations of $3\,\Delta$ in the $x$-space (or $s=3$ in the scaled separation variable $s$); we don't need all neighbors.
Beyond separation $s=3$, the kernel and its first two derivatives exactly vanish everywhere.
Density estimation with this kernel can be performed with only neighbors out to $s=3$.
We obtain the list of $s\leq 3$ neighbors exactly using the scipy \citep{scipy} ckdtree implementation of the kd-tree, which has scales like $\ln N$ rather than $N$ for problems like ours (where the bandwidth is smaller than the distribution of points).
This can potentially speed up the density estimation enormously relative to methods that sum over all points.


\subsection{1-D gaps: valleys in the density field}

\revision{Critical points are points where the derivatives (slopes) of the density estimate are zero  (i.e. they can be a minimum, a maximum or a saddle). Since a gap will necessarily contain either a saddle point or a minimum point, critical points could be used to pinpoint gaps-region in a local way. 

First, we present an empirical way} to approximate critical points in a given dataset (or point-cloud) in practice (although other methods could be used, e.g. as in \cite{rosen2021using}). 
\\Then, we propose a statistic that can be used to rank critical points in order to select the ones that lie in the gaps. We show on our application that this statistic is a sound proxy to select  critical points lying in gaps in the distribution. 
\\From these points, we then show how using the gradient of the density field can trace and highlight specific types of gaps, flowing from the saddle points, in the forms of ``valleys' or ``streak'' in the density flow. This approach thus allows to focus on `1-dimensional' gaps. We provide examples on our application to illustrate in which setups this might prove useful, e.g. for easier downstream analysis or visualization, by applying this ``valley-finder'' on 2D-slices from our 3D datasets.





\subsubsection{Approximating critical points}

To approximate the critical points within the boundary $S$ of $\mathbb{R}^d$ for a given estimator $p_{\Delta}$, we propose the following methodology: 
\begin{enumerate}
    \item Create a grid of points $G$ within the boundaries $S$, spaced by $\Delta$, the bandwidth for the density estimator.
    \item For each point $x \in G$, optimize to minimize the squared gradient of the density estimate. \revision{We use BFGS algorithm provided in \textit{scipy} optimization library}. 
    \item Group the resulting critical points so that points close to each other (e.g. within a distance function of the bandwidth $\Delta$) are aggregated. In the following, we group together points that are closer than $0.01 \Delta^2$ and \revision{summarize them as a single point being the mean of the group of points.}
\end{enumerate}

This process allows us to empirically approximate the set of relevant critical points (i.e. minimum, maximum and saddles) in the density field, for a given bandwidth. \revision{However, we note that this method might be sub-optimal in terms of computational efficiency especially as the number of dimensions increases. We propose to use it here for sake of simplicity and practicality for our use-case. Other methods can be used, e.g. relying on contour trees as in \cite{rosen2021using}.}

\begin{figure}
     \gridline{ \fig{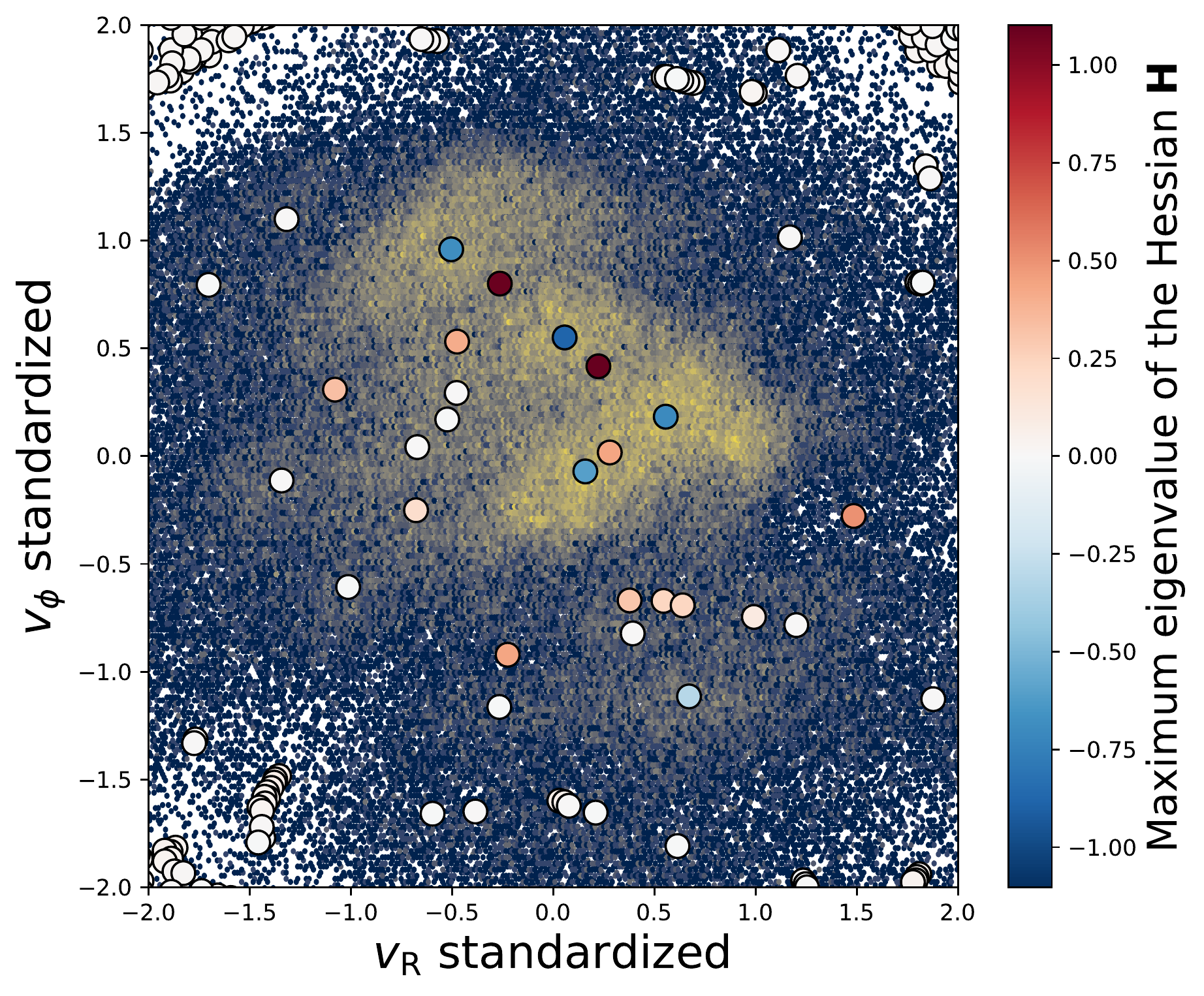}{0.5\textwidth}{(a)}
                \fig{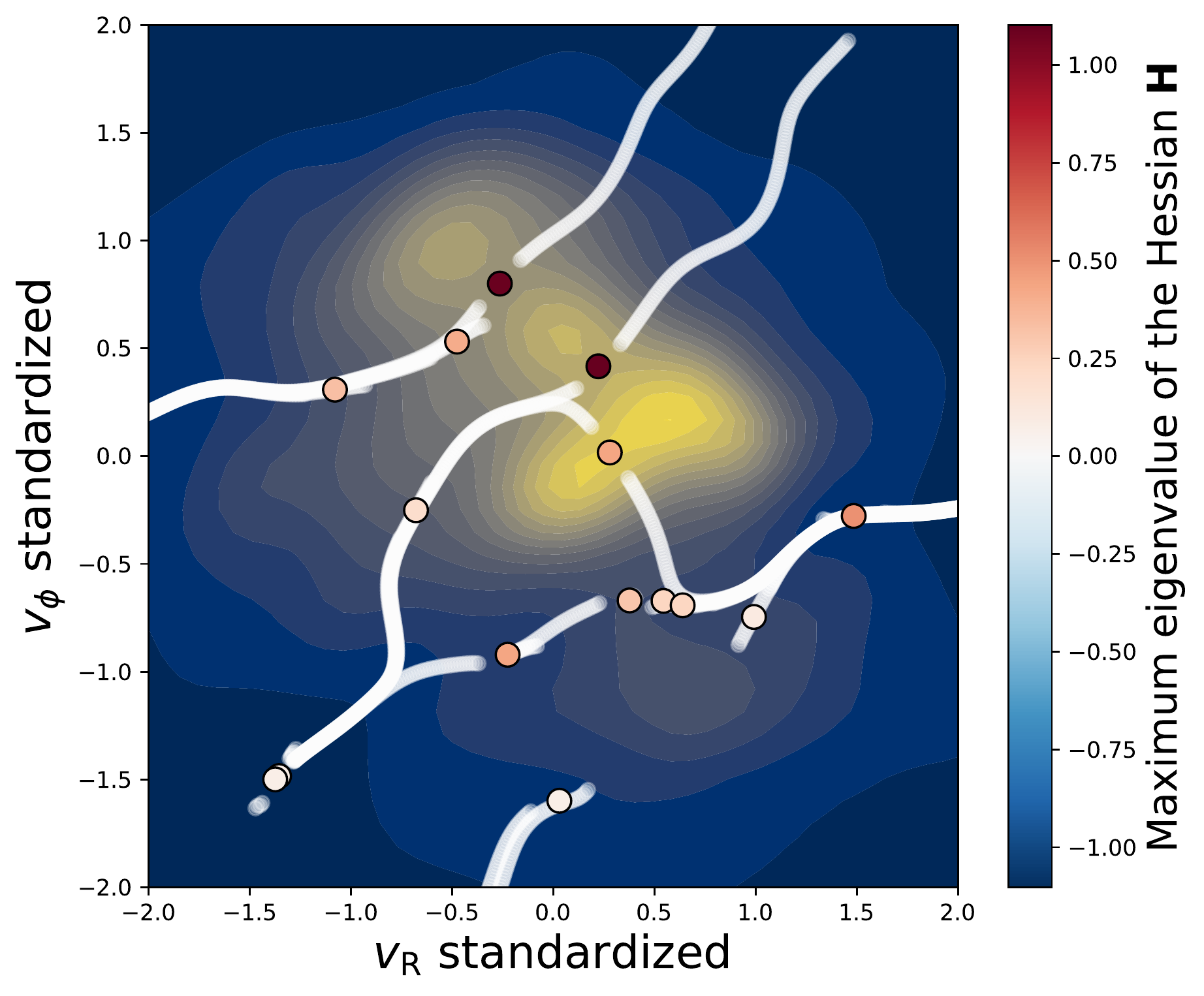}{0.5\textwidth}{(b)}
            }
        \caption{Left: Critical points found with a density estimator of bandwidth $\Delta = 0.15$, circled in black and colored by the value of the maximum eigenvalue of the Hessian of their density, for 2-dimensional dataset $D_R$. Critical points with negative value (blue) corresponds to maximum, while positive points corresponds to local minimum and saddle. Points with values closest to zero lie mostyl in flat region of the density space. 
        Right: Selection of N=15 best critical points based on the value of the maximum eigenvalue of the Hessian of their density, with their `paths' (following the gradient direction) traced in white. \label{fig:criticals}}
        
\end{figure}

\subsubsection{Ranking critical points}

We now need a way to rank the critical points in order of interest. First, it is relevant to point out that the nature of a critical point (i.e. if it is maximum, minimum or saddle) can be determined through its \textit{index}, i.e. the number of negative eigenvalues of the Hessian matrix $\mathbf{H}$ (second derivative) of the density estimate at that point, since the eigenvalues of the Hessian give us indications on the direction of the curvature of the density field. A maximum point will have an index of $d$ (number of dimensions of the data space), while a minimum point will have an index of 0. Critical points with indexes above 0 and below $d$ are saddle points. This can potentially be used to reduce the set of critical points in the first place.

\revision{The statistic presented in this paper is undefined exactly at critical points, and thus can not be used as a ``score'' to rank the critical points. However, the maximum eigenvalue of the Hessian matrix is a good proxy criterion for our task, as it will rank higher the critical points that have the biggest positive curvature in the density field.} While in theory this criterion selection could end up selecting critical points that are not corresponding to gaps by our definition (e.g. if the density field is in the form of a cliff with a perfectly flat bottom, a minimum critical point could land at the bottom there and have a high score), we observe that it is in practice a sensible choice, as the critical points `at cliffs' will move far enough away to have a lower maximum eigenvalue of the Hessian. 

As an illustration, we compute the critical points estimated on the 2-D dataset $D_v$ described in Section \ref{sec:data}, with a density estimator with a bandwidth $\Delta$ of 0.15. Figure \ref{fig:criticals}a shows the resulting critical points colored by the value of the maximum eigenvalue of the Hessian at that location. We see that the critical points with the lowest value corresponds to maximum, while the critical points with the highest values lie on saddle or local-minimum regions. The critical points with values closer to zero tend to be in flatter, outer-skirt regions.

\subsection{Tracing `1-D' valley-gaps}

We now proceed to use the ``best'' candidates critical points (i.e. the ones with the highest scores) as a `starting point' to highlight gaps: indeed, while those points identify the local minimum point of the basin formed by a gap, they do not cover nor characterize the entire gap `region' (or in this instance, `valley') per se. However, we can see a saddle point as the `origin' of a valley gap, from which the gradient of the density can either go uphill (in the direction of the Hessian's eigenvectors with positive eigenvalues) towards higher density regions, or downhill (in the direction of the Hessian's eigenvectors with negative eigenvalues) towards lower density regions. We propose to use these geometrical properties of the density field to trace the valley from a saddle point $x$ in practice as follow:

\begin{enumerate}
    \item Starting from a critical point $x$, take a small (e.g. a fraction of the bandwidth $0.1 \Delta$) step away from $x$ in the direction of the Hessian's eigenvector associated to the smallest Hessian's eigenvalue.
    \item Descend `downhill' following the gradient of the density field in small steps of  e.g. a fraction of the bandwidth $0.1 \Delta$ until a low gradient is reached or until reaching the boundary $S$.
\end{enumerate}
Figure \ref{fig:criticals}b illustrates the paths we obtain with this process on the 2D dataset $D_v$, with a density estimator with a bandwidth $\Delta=0.15$, selecting the best 15 critical points using the maximum eigenvalue of their Hessian. We can see that those paths neatly trace the visible valleys in the density distribution. Some of the selected critical points end up being connected to each other.



\begin{figure} 
    \gridline{
        \fig{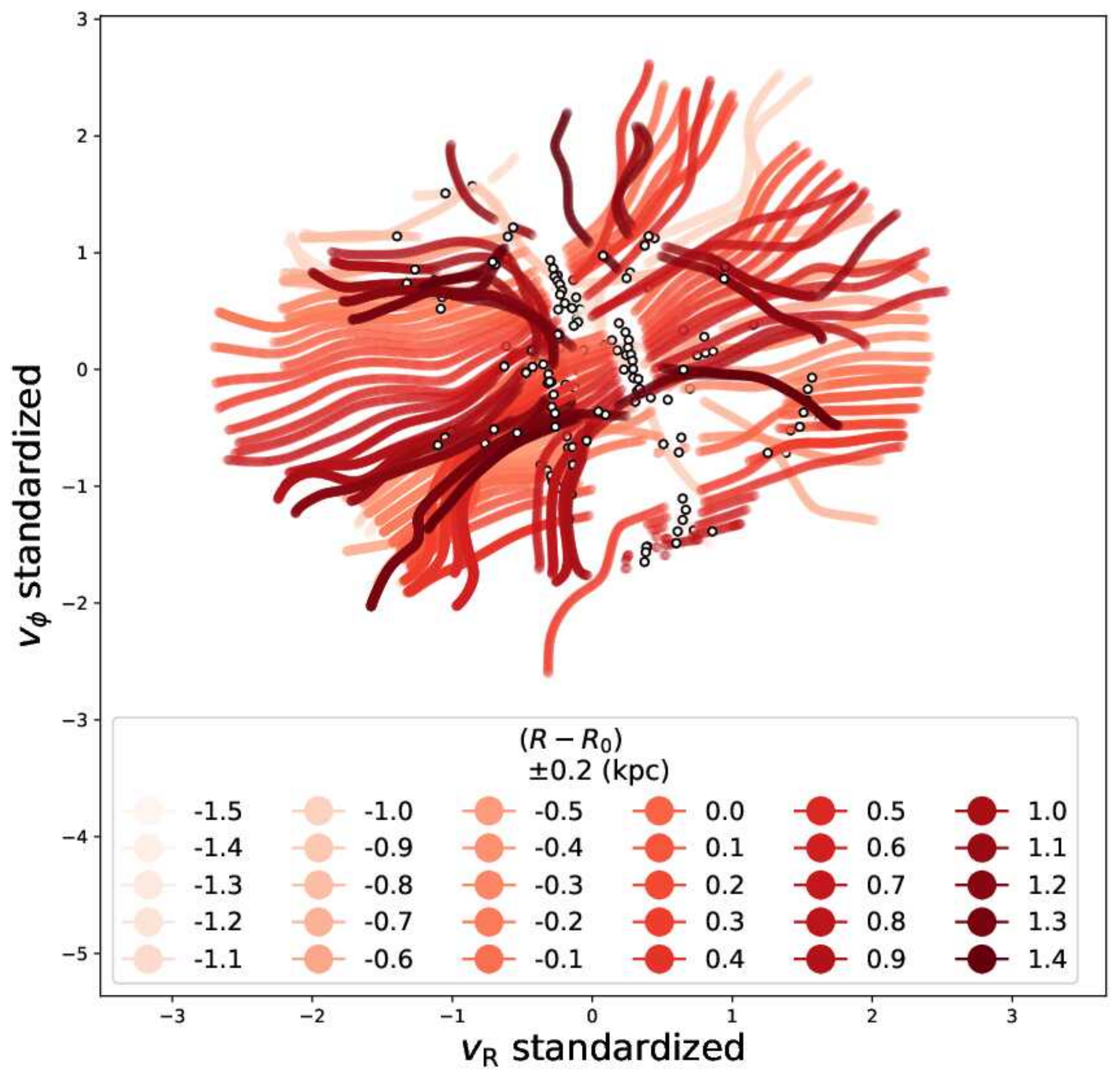}{.5\textwidth}{(a)}
        \fig{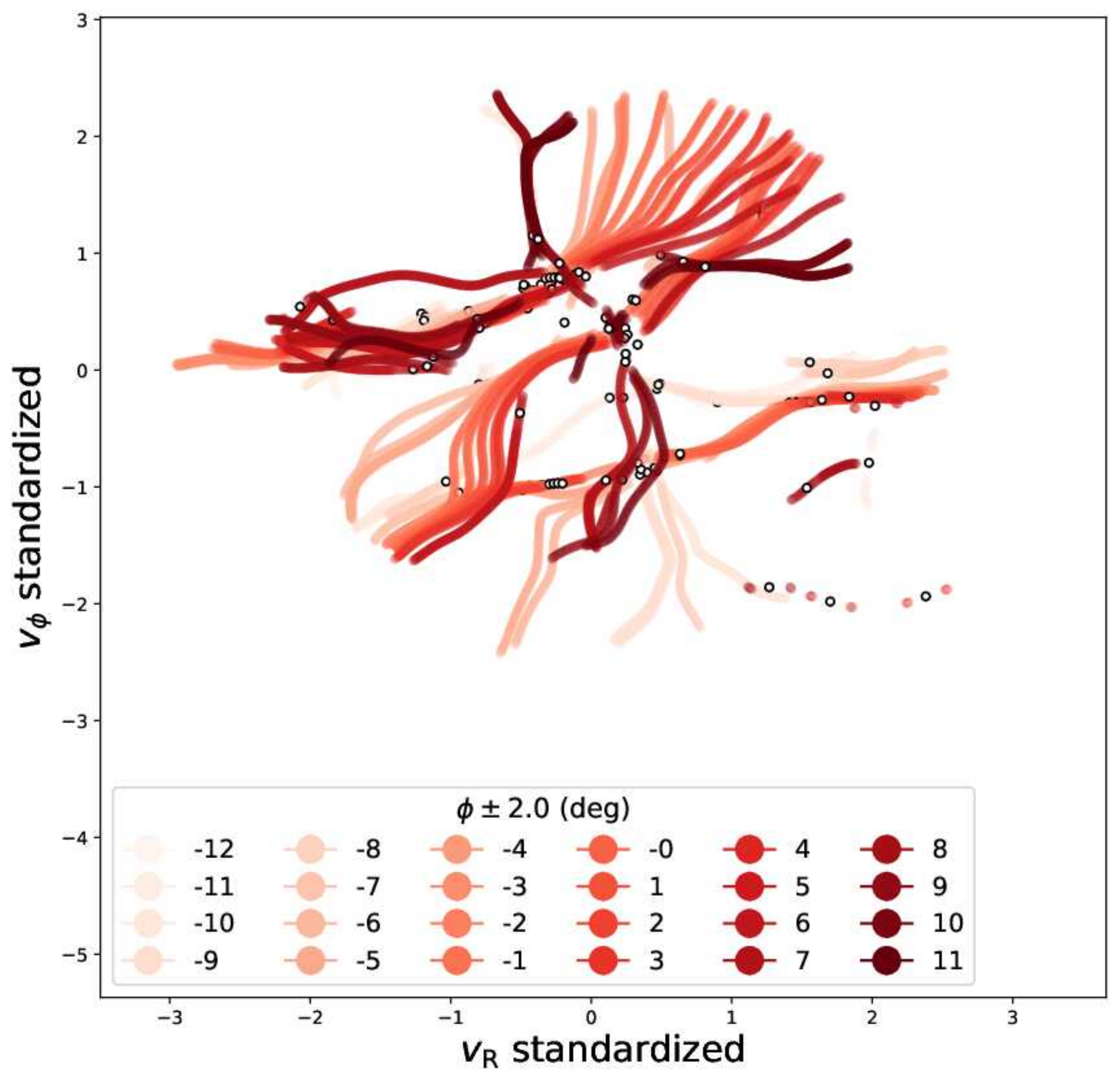}{.5\textwidth}{(b)}
    }
    \caption{Best 5 critical points and their associated paths for each 2D-`slice' in respectively $D_R$ (left pannel) and $D_\phi$ (right figure), colored by their respective $R$ and $\phi$ window value, with density estimators of bandwidth 0.15. \label{fig:slicing}}
\end{figure}

\revision{While tracing the ``valleys'' in this way restricts us to 1-D gaps, it can provide an interesting method for specific use-case, if one has knowledge of the topology of the gaps of interest. Additionally, it might make the visualization and characterization of the properties of the gaps easier in higher dimension. We illustrate a different experimental protocol to analyze the gaps in a 3-dimensional dataset, where we ``slice'' the original dataset into 2-D chunks. We then compute the critical points, rank and select the K ``best'' critical points, and finally compute their ``paths'' on slices (subset) of the original data instead, resuming to a 2-D setting for each run.

Figure \ref{fig:slicing} illustrates this methodology on our original 3-dimensional dataset $D_R$ (Fig \ref{fig:slicing}a) and $D_\phi$ (Fig \ref{fig:slicing}b). The datasets were respectively sliced into 30 and 24 slices creating 2-D datasets with overlapping windows of similar size ($0.2$ kpc in $R-R_0$ for $D_R$, and 0.035 rad in $\phi$ for $D_\phi$). The bandwidth used for the density estimator was $0.15$. For each slice, we keep the best 5 critical points. This methodology highlights again the change in gaps location as a function of $R$ and $\phi$. Additionally, we can visually see the critical points that seem to belong to the same gaps. The inclination and rate of change also look different for each gap.  Another interesting aspect of this methodology, compared to the selection using our \pihpi criterion in Figure \ref{fig:pihpi_3d}, is that the ``paths'' can extend to region that would not be selected using a \pihpi-based cut (note the axis range of Figure \ref{fig:slicing}). This approach might provide an easier path to properly analyze the rate and nature of the change, by characterizing each path individually, instead of the entire 3-D structure or region selected through gridding.

We defer a deeper analysis of the characterization of the gaps and their links to physical origins to future works.}

\end{document}